\documentclass{iucrjournals}

\usepackage{lineno}
\usepackage{float}
\usepackage{color} 
\usepackage{comment}
\usepackage{
graphicx}
\usepackage{adjustbox}
\usepackage{caption}
\usepackage{subcaption}
\usepackage{isotope}
\usepackage{textgreek}
\usepackage{soul}

\usepackage{url}

\title{Evaluation of a Full Field Fluorescence Imager with Synchrotron Radiation}

\author[a]{Piotr Maj\IUCrCemaillink{pmaj@bnl.gov}\IUCrOrcidlink{0000-0002-1059-6376}}%
\author[a]{Grzegorz W. Deptuch\IUCrCemaillink{gdeptuch@bnl.gov}\IUCrOrcidlink{0000-0003-1703-6758}}%
\author[a]{Dominik S. Gorni\IUCrCemaillink{dgorni@bnl.gov}\IUCrOrcidlink{0000-0001-9613-7180}}%

\author[a]{Giovanni Pinaroli\IUCrCemaillink{gpinaroli@bnl.gov}\IUCrOrcidlink{0000-0001-6739-0339}}%

\author[a]{Gabriella A. Carini\IUCrCemaillink{carini@bnl.gov}\IUCrOrcidlink{0000-0003-2642-5106}}%

\author[b]{David P. Siddons\IUCrCemaillink{siddons@bnl.gov}\IUCrOrcidlink{0000-0003-1482-9867}}%
\author[b]{Ryan Tappero\IUCrCemaillink{rtappero@bnl.gov}\IUCrOrcidlink{0000-0002-3560-2461}}%
\author[a]{Soumyajit Mandal\IUCrCemaillink{smandal@bnl.gov}\IUCrOrcidlink{0000-0001-9070-2337}}%
\author[a]{Donald Pinelli\IUCrCemaillink{pinelli@bnl.gov}\IUCrOrcidlink{0009-0005-7090-3613}}%
\author[a]{Timothy Kersten\IUCrCemaillink{tkersten@bnl.gov}\IUCrOrcidlink{0009-0003-4217-5503}}%

\author[a]{Nicholas~St.\,John\IUCrCemaillink{nstjohn@bnl.gov}\IUCrOrcidlink{0000-0002-8758-1877}}%

\author[b]{Abdul K. Rumaiz\IUCrCemaillink{rumaiz@bnl.gov}\IUCrOrcidlink{0000-0002-9907-299X}}%

\author[b]{Anthony Kuczewski\IUCrCemaillink{kuczewski@bnl.gov}\IUCrOrcidlink{xxxx-xxxx-xxxx-xxxx}}%

\affil[a]{Instrumentation Department, Brookhaven National Laboratory, PO Box 5000, Upton 11973-5000 NY, USA}
\affil[b]{NSLS-II, Brookhaven National Laboratory, PO Box 5000, Upton 11973-5000 NY, USA}
            
\begin{document} 
\maketitle 

\begin{synopsis}
A new event-driven, energy-resolving ASIC for full-field X-ray fluorescence imaging is presented. It was validated on the NSLS-II beamline, confirming its frame-less mode of operation, and providing energy measurements in a two-dimensionally segmented detector translating to spectroscopic capabilities achieved without mechanical scanning. Leveraging advanced on-chip spectral acquisition alongside autonomous readout, the system substantially broadens the potential for real-time elemental mapping in biological, environmental, and materials research.
\end{synopsis}

\begin{abstract}
The design and evaluation on the NSLS-II beamline of the 3FI application specific integrated circuit (ASIC) bump-bonded to a simply, planar, two-dimensionally segmented silicon sensor is presented. The ASIC was developed for Full-Field Fluorescence spectral X-ray Imaging (3FI). It is a small-scale prototype that features a square array of 32 $\times$ 32 pixels and the size of the pixels is 100\,\textmu{m} $\times$ 100\,\textmu{m}. The ASIC was implemented in a 65\,nm CMOS integrated circuit fabrication process. Each pixel incorporates a charge-sensitive amplifier, shaping filter, discriminator, peak detector, and sample-and-hold circuit, allowing detection of events and storing signal amplitudes. The system operates in an event-driven readout mode, outputting analog values for threshold-triggered events, allowing high-speed multi-element X-ray fluorescence imaging. 

At power consumption of 200\,\textmu{W} per pixel, consisting almost uniquely of power dissipated in analog blocks, 308 eV full width at half maximum (FWHM) energy resolution at 8.04\,keV, that corresponds to 30\,e\textsuperscript{-} rms equivalent noise charge (ENC) and 138 eV FWHM energy resolution at 3.69\,keV (16\,e\textsuperscript{-} rms ENC) were obtained, for Cu and Ca K\textsubscript{\textalpha} lines, respectively. Each pixel operates independently in the 3FI ASIC. Therefore the detector enables in situ trace element microanalysis in biological and environmental research. Its architecture addresses limitations of X-ray Fluorescence Microscopy (XFM), typically requiring mechanical scanning, by offering frame-lees data acquisition, translating to high-throughput operation.

The 3FI ASIC is suitable for example for studies of nutrient cycling in the (mycor)rhizosphere, microbial redox processes, and genotype-phenotype correlations in bio-energy crops. Additional performances, such as enhanced spatial resolution can be further improved with coded-aperture and Wolt, extending the use to environmental, biomedical, and material science studies. The successful test of the small-scale prototype encourager scaling up the design to large-area $>$10\,kpixels.
\end{abstract}


\section{Introduction}

Hybrid pixel detectors have been widely used in X-ray imaging and scientific experiments for over thirty years. Their architecture, which allows the sensor material to be selected independently from the readout electronics, makes them particularly well-suited for applications requiring high spatial resolution, wide dynamic range, and good sensitivity. When sensor and the readout integrated circuit (ROIC) are fabricated separately, different semiconductor technologies can be used aiming at the optimized performance for the specific application and energy range.

Most of the renowned detectors used in synchrotron applications are hybrid pixelated detectors. Notable examples include systems developed by PSI (Jungfrau ~\cite{Mozzanica02112018}, CERN (Medipix and Timepix~\cite{ballabriga2013medipix3rx}, ~\cite{YOUSEF2017639}), Dectris (Pilatus, Eiger, and others~\cite{Broennimann2006}, ~\cite{Dinapoli2011}, ~\cite{dinapoli2013eiger}, ~\cite{Donath:gy5047}), Rigaku (XSPA~\cite{zhang2016submillisecond}), DESY (Labmda ~\cite{Pennicard_2014}), and SLAC (ePIX~\cite{vanDriel2020ePix10k}). These commercially available detectors are optimized for different performance parameters such as energy resolution, pixel size, maximum count rate, and detector uniformity-defined in part by inter-pixel and inter-detector energy dispersion.

\begin{figure}[ht] 
\begin{center}
\includegraphics[width=0.66\textwidth]{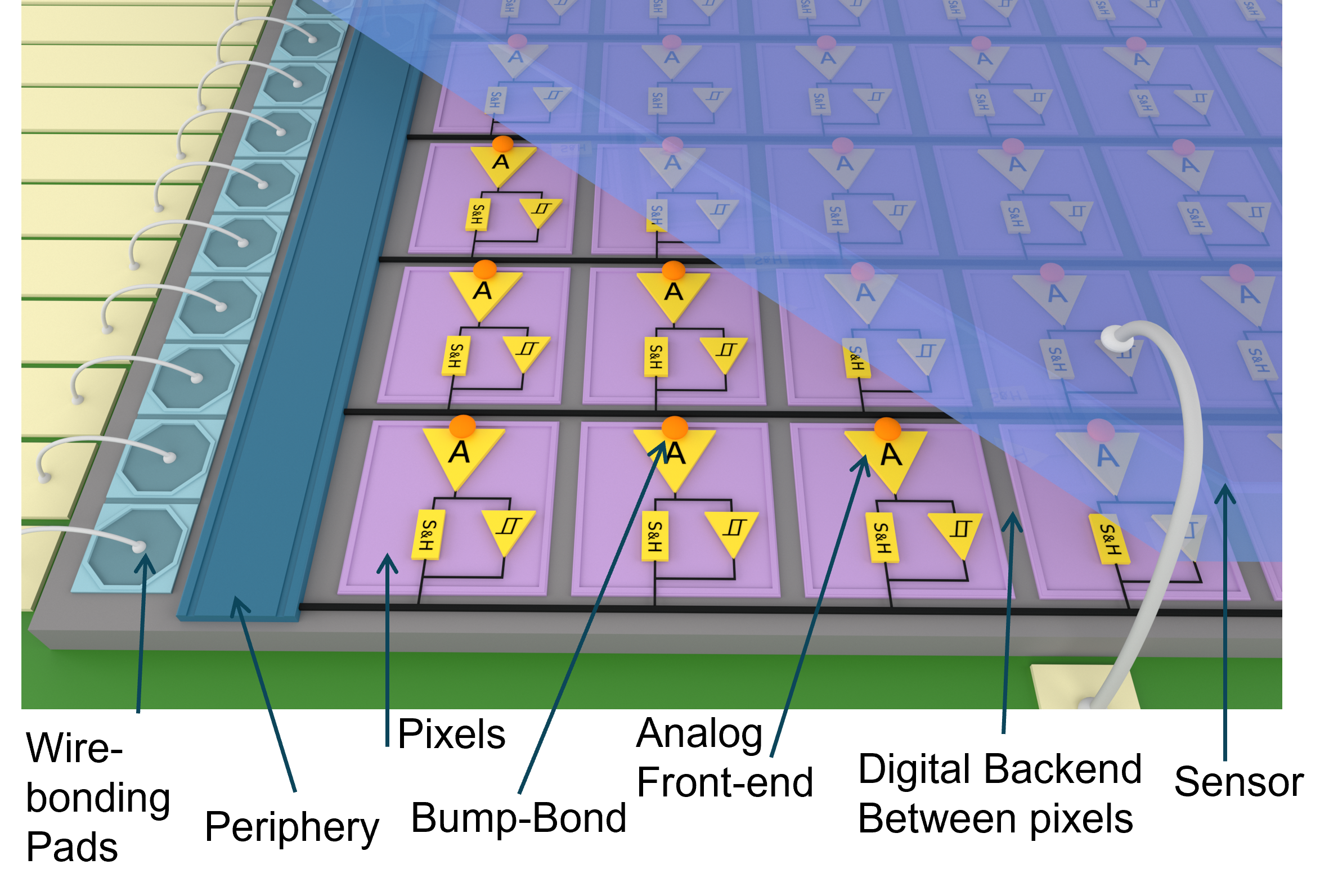} 
\end{center}
\caption{A hybrid pixel detector - idea of operation.} 
\label{fig:HybriDetIdea}
\end{figure}

The operating principle of hybrid pixel detectors is illustrated in Fig.~\ref{fig:HybriDetIdea}. A photon interacting with the sensor liberates charge in its active volume, producing a very short and low-amplitude current pulse that serves as the input to the readout electronics connected to the sensor with bump-bonding technique. The ROIC ultimately defines the detector’s functionality and performance. For single-photon processing, each pixel channel typically contains a charge-sensitive amplifier (CSA) and a shaping stage, which together determine noise characteristics and signal processing speed. Once the signal is shaped and filtered, it can, for example, be compared to a threshold and counted, enabling operation at high photon fluxes. This is the basis for the vast majority of commercial systems.

More demanding applications require not only for counting individual events but also for measuring their amplitude and its value readout without noticeable system dead-time. For over two decades, researchers are trying to enable per-pixel spectroscopy~\cite{Deptuch2024}, but limitations in technology nodes and ROIC backend architectures have hindered practical implementations. Most existing efforts can discriminate between a limited number of energy levels using multiple thresholds but do not provide direct photon energy information.

Recent developments at Brookhaven National Laboratory (BNL) have led to a new generation of energy-discriminating readout application specific integrated circuits (ASICs), capable of capturing the amplitude of each detected photon. This advancement is made possible by a novel, in-house developed event-based readout paradigm known as EDWARD~\cite{Gorni_2022}, which enables immediate output of digital address and analog signal data for each event. This type of readout architecture is essential for advanced X-ray imaging modalities such as X-ray Fluorescence Microscopy (XFM) and the emerging Full-Field Fluorescence spectral X-ray Imaging (3FI or FFFI).

\subsection{Motivation for Full-Field X-ray Fluorescence Imaging}

X-ray Fluorescence Microscopy is indispensable in mapping elemental compositions and distributions in biological, environmental, and materials science samples ~\cite{Copeland-Hardin2023}. Traditional scanning-based XFM methods, although capable of delivering excellent spatial resolution, suffer from long acquisition times because fluorescence data must be collected in steps of many points of exposure. This limitation inhibits efficient studies of dynamic biochemical processes, such as elemental transport, oxidation-state changes, and chemical transformations in living systems~\cite{Kourousias2020}.

Full-Field X-ray Fluorescence Imaging addresses these challenges by simultaneously capturing fluorescence signals from an entire field of view, enabling near real-time elemental imaging. In contrast to scanning systems, Full-Field X-ray Fluorescence Imaging requires segmentation of a detector that can be achieved in arge-area pixelated detectors~\cite{Romano2014}. This capability is especially valuable for studying biological interfaces such as plant–soil interactions, microbial redox processes, and trace element dynamics—situations where elemental distributions and oxidation states can evolve rapidly~\cite{Brinza2014}.

This detector, together with appropriate optics, will provide elemental maps in 2D without mechanically scanning the sample, and 3D, elementally-resolved tomograms with a single translation~\cite{Siddons2020}. This will provide an important speedup of such experiments, particularly for biological samples.

\subsection{Synchrotron-Based Evaluation of 3FI and Paper Structure}

To demonstrate the principle of Full-Field X-ray Fluorescence Imaging the 3FI ASIC, a prototype detector was designed, fabricated and then hybridized with a sensor to  evaluate at NSLS-II. Its ability to capture trace element distributions was assessed using standard fluorescent secondary radiation from reference samples.

This paper presents the design, implementation, and experimental validation of a photon-spectrometric 3FI-based small-scale prototype detector. Section \ref{Sec:3FI_ASIC} introduces the architecture of the ASIC prototype. Section \ref{Sec:3FI_detector} describes the overall detector system and integration strategy. Section \ref{Sec:Measurements} presents measurement results, focusing on the results achieved in  experiments using synchrotron beamline and highlighting the detector’s energy resolution and event-driven capabilities.

\section{The 3FI ASIC for Color X-ray Imaging}\label{Sec:3FI_ASIC}

3FI stands for Full-Field Fluorescence Imager. It is an Application-Specific Integrated Circuit, also referred to as a Readout Integrated Circuit (ROIC), suitable for reading out a pixelated semiconductor sensor  that was developed at Brookhaven National Laboratory using a 65~nm CMOS process node. The 3FI ASIC is specifically designed for near real-time trace element microanalysis in complex biological systems.

\begin{figure}[ht] %
\begin{center}
\includegraphics[width=0.5\textwidth]{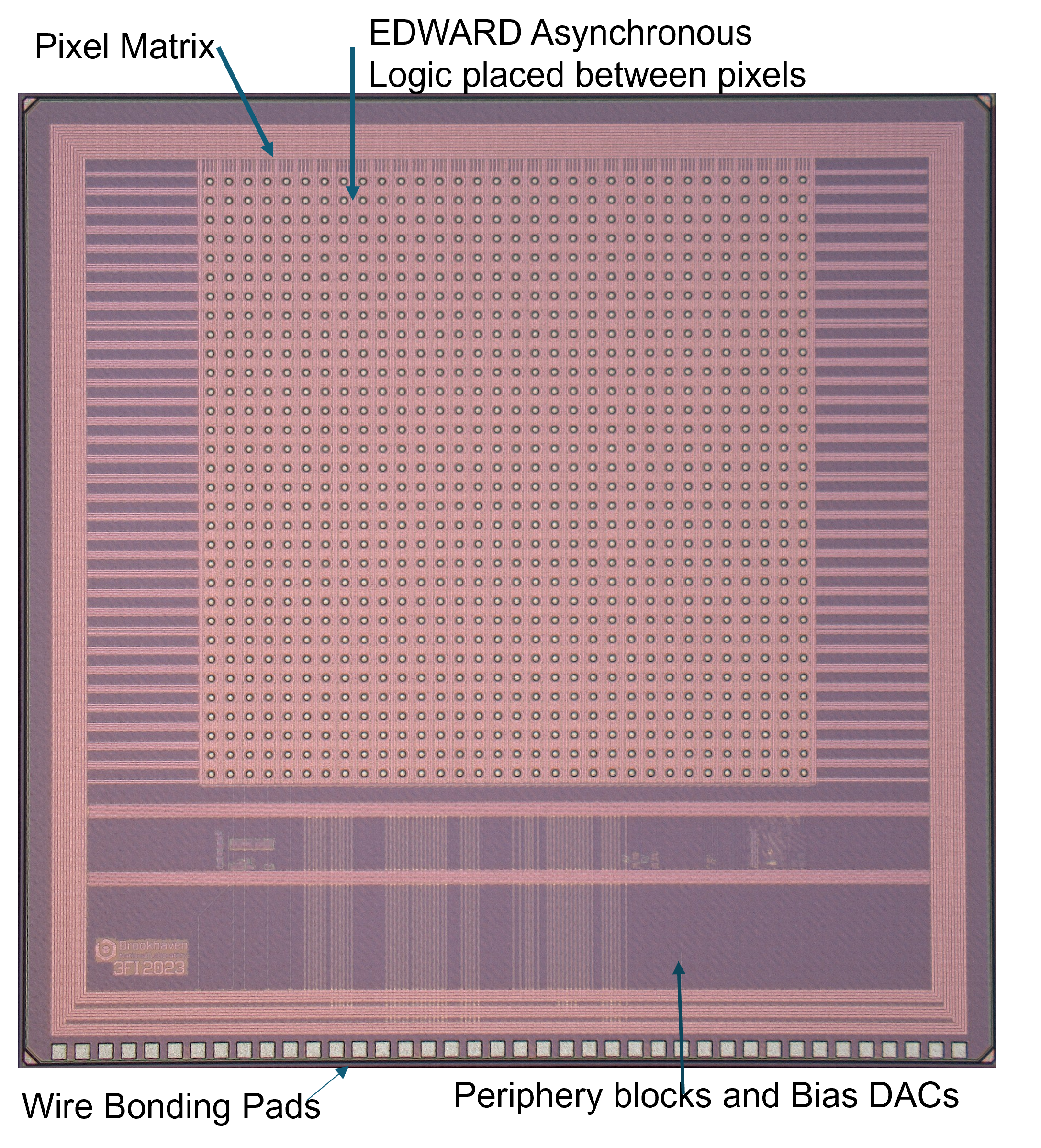} 
\end{center}
\caption{Photograph of the 3FI ASIC with visible wire-bonding pads on placed on one side supporting 3-side bootable detector designs, periphery blocks for chip control, pixels and EDWARD logic layout in between them.} 
\label{fig:3FIPhoto}
\end{figure}

A full-scale chip aims at a matrix of 256~×~256 pixels; however, the current prototype comprises an array of 32~×~32 square pixels. The pitch of a pixel is 100~\textmu{m}. The pixel matrix operation is supported by peripheral circuitry providing among others proper biasing, IC control and a temperature sensor. A photograph of the prototype chip is shown in Fig.~\ref{fig:3FIPhoto}.

\begin{figure}[ht]
\begin{center}
\includegraphics[width=0.9\textwidth]{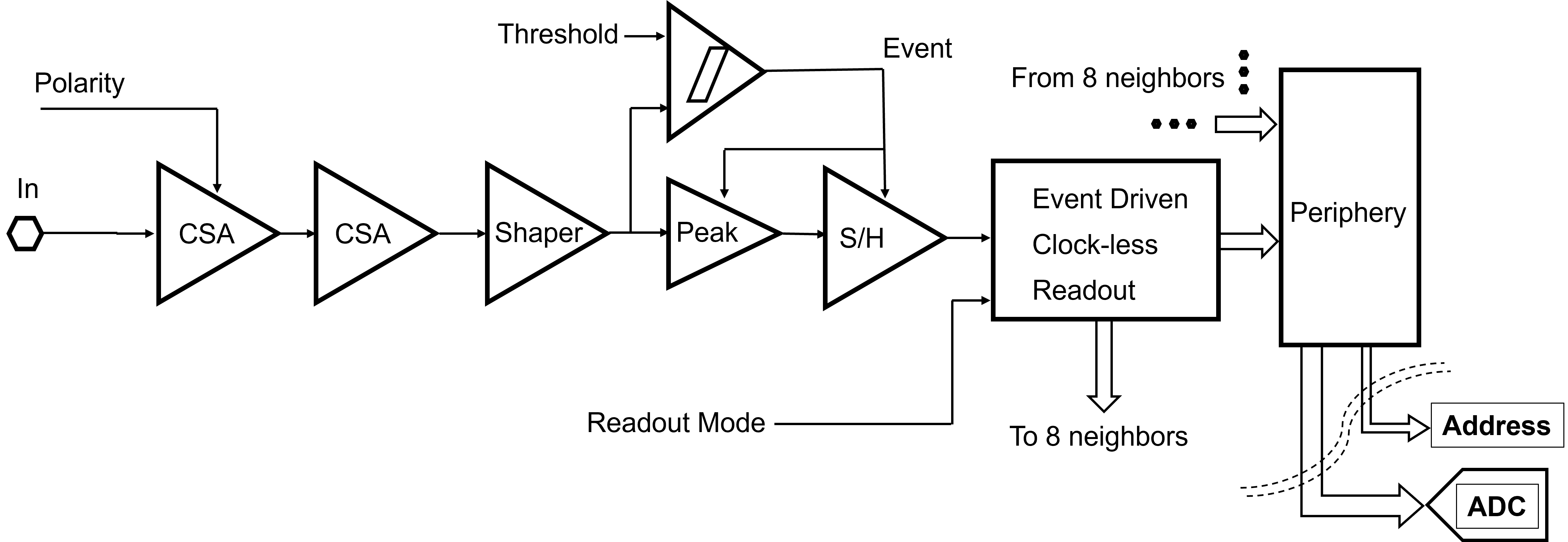} 
\end{center}
\caption{Simplified schematic diagram of a single channel in the 3FI ASIC.}
\label{fig:3FIChannel}
\end{figure}

A simplified block diagram of a single 3FI analog front-end channel is shown in Fig.~\ref{fig:3FIChannel}. Each channel consists of a two-stage charge-sensitive amplifier (CSA), a third-order pulse-shaping filter, a discriminator with offset correction, a peak detector, and a bank of nine sample-and-hold (S\&H) circuits. This configuration enables the readout of the active pixel and its eight immediate neighbors. An active pixel is defined by the discriminator output, that triggers an event for the channel. An event-driven, asynchronous logic manages the event address and associated analog value readout and routes both analog signal and digital address through the peripheral circuitry to the analog and digital output buffers respectively.

 Input charge processing is optimized for low-noise operation by employing a high-gain, two-stage amplifier followed by a third-order, semi-gaussian pulse filter. When an input pulse exceeds the threshold level, the discriminator triggers a valid event, prompting the peak detector to capture the maximum signal value. This value is stored in the S\&H array, which simultaneously latches the values of the central pixel and its eight neighbors. These latched values are then read out via the EDWARD asynchronous logic.

The 3FI ASIC outputs data through two channels: the first channels is for analog values and features a differential analog output buffer that transmits the in-pixel-stored values of signal amplitudes. The second channel is digital and it contains a high-speed serializer, that outputs digital addresses of occurrences of radiation interaction events using the LVDS-compatible differential signaling mode. Both output paths are fully configurable, allowing fine control over the digital waveform characteristics~\cite{St.John_2023}.

\begin{figure}[ht] %
\begin{center}
\includegraphics[width=0.7\textwidth]{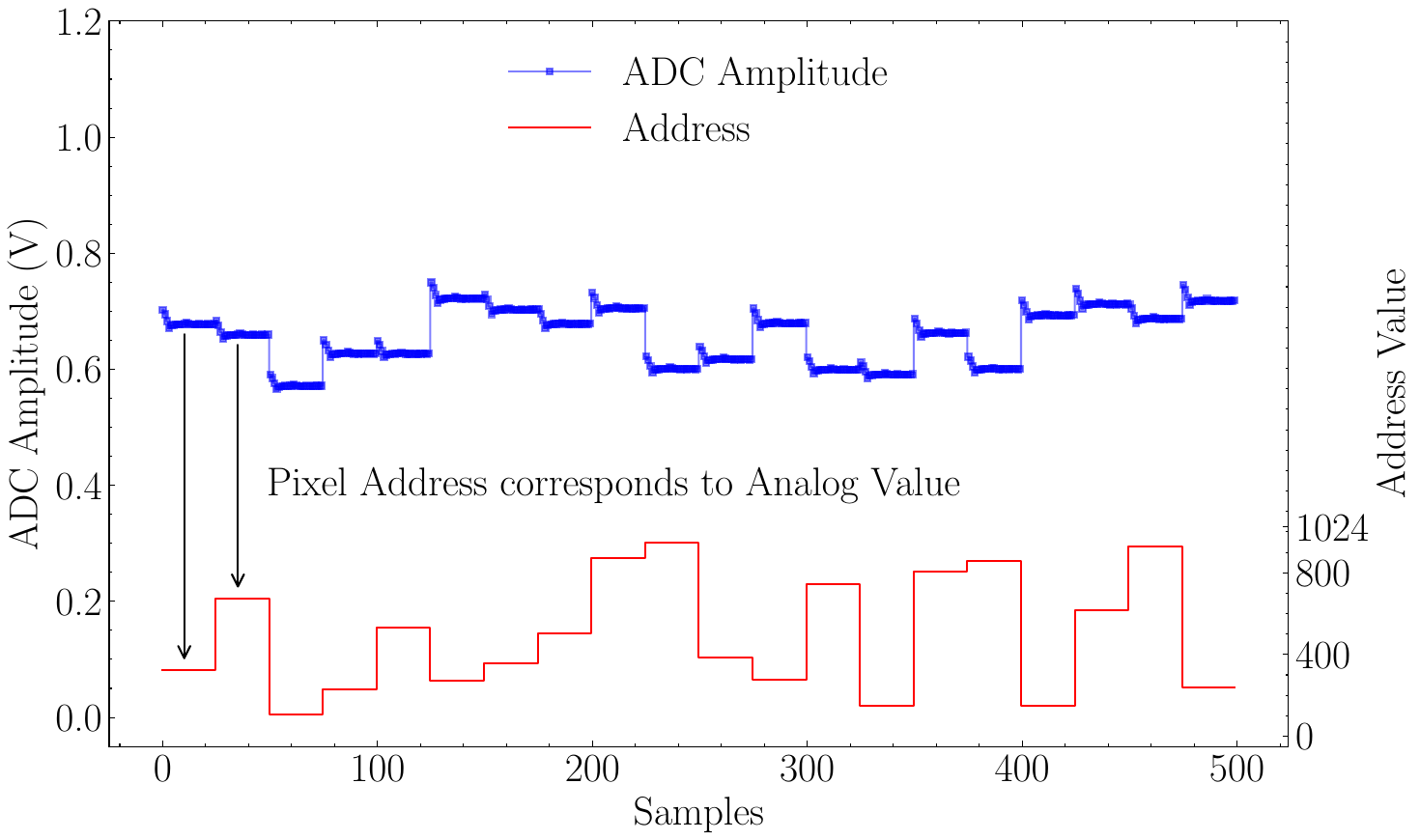} 
\end{center}
\caption{3FI Readout modes: single-pixel readout mode, where the upper plot is assigned to the left scale presenting the measured pulse amplitude in voltage, while the bottom plot presents the corresponding event address scaled to the right-side Y axis.} 
\label{fig:3FIReadMode1}
\end{figure}

\begin{figure}[ht] %
\begin{center}
\includegraphics[width=0.7\textwidth]{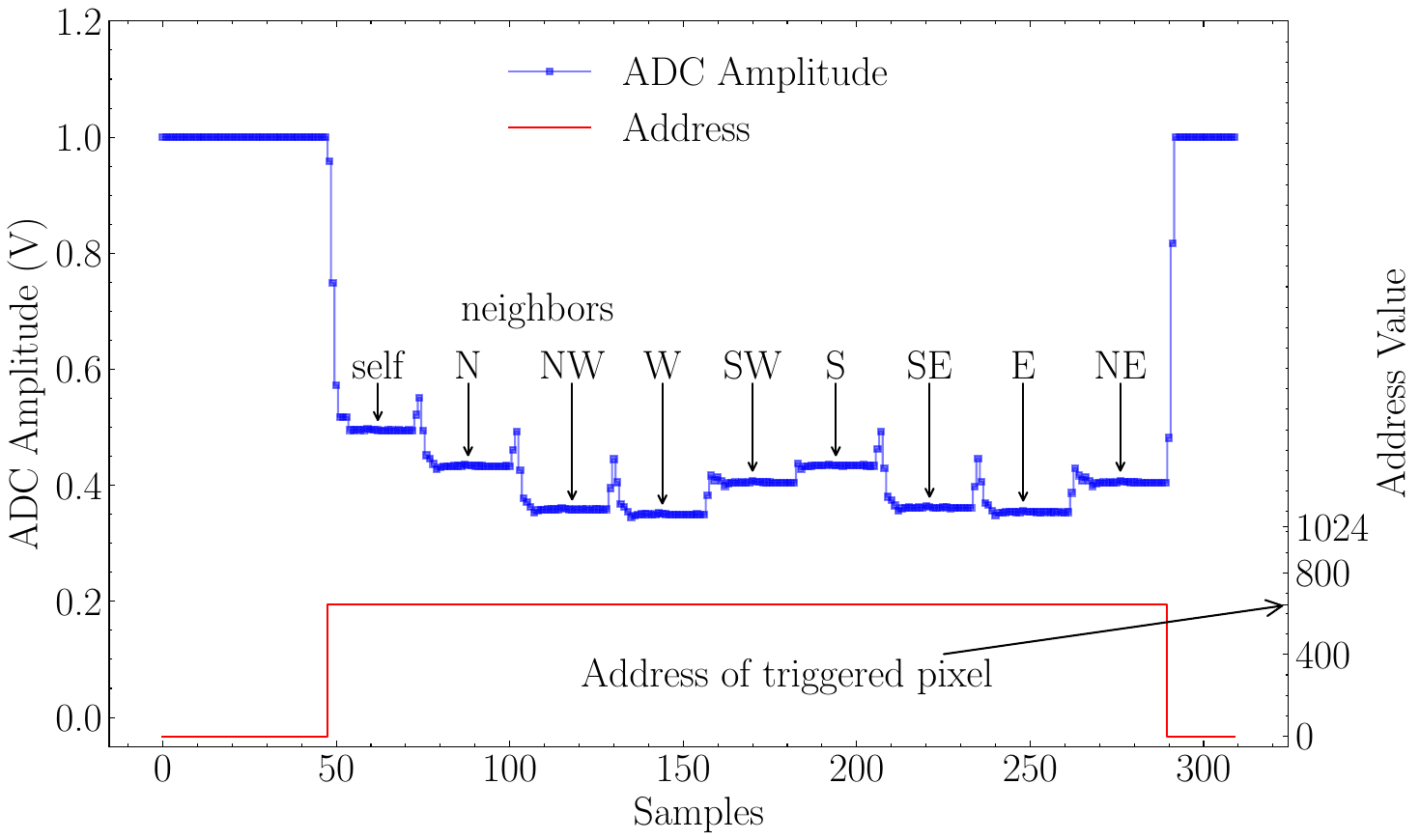} 
\end{center}
\caption{3FI Readout modes: charge-sharing compensation readout mode, where the upper plot is assigned to the left scale presenting the measured pulse amplitude in voltage, while the bottom plot presents the corresponding event address scaled to the right-side Y axis. The single event from the pixel address 648 triggers the readout of a central pixel together with all neighbors marked with geographical shortcuts.} 
\label{fig:3FIReadMode2}
\end{figure}

The 3FI ASIC is designed for zero-dead-time operation, ensuring that the readout of an active event does not interfere with the activity of the remaining channels. It can operate in two readout modes, as illustrated in Fig.~\ref{fig:3FIReadMode1} and Fig.~\ref{fig:3FIReadMode2}. In the first, single-pixel readout mode, a discriminator trigger causes latching the values of all nine neighboring pixels, but readout of only the central pixel leaving the remaining 8 values unread. In the second mode, referred to as the charge-sharing compensation readout mode, a single trigger allows reading out of all nine neighboring pixels' values. These signals can be then processed outside the ASIC, within the detector's embedded system, to reconstruct the total photon energy.

This compensation mechanism implemented in the 3FI addresses the charge-sharing effect, a significant contributor to energy resolution degradation. The impact of charge-sharing on overall system performance depends on several factors, including X-ray photon energy, sensor material (with more pronounced effects in CZT), sensor thickness, pixel dimensions, and applied bias voltage ~\cite{Maj_2015, 6527518, krzyzanowska2017characterization}. While the single-pixel readout mode offers higher throughput, the charge-sharing compensation mode is designed to improve energy resolution by an off-chip correction for charge shared between adjacent pixels.

\section{The Detector Design}\label{Sec:3FI_detector}

\begin{figure}[ht]
\begin{center}
\includegraphics[width=0.5\textwidth]{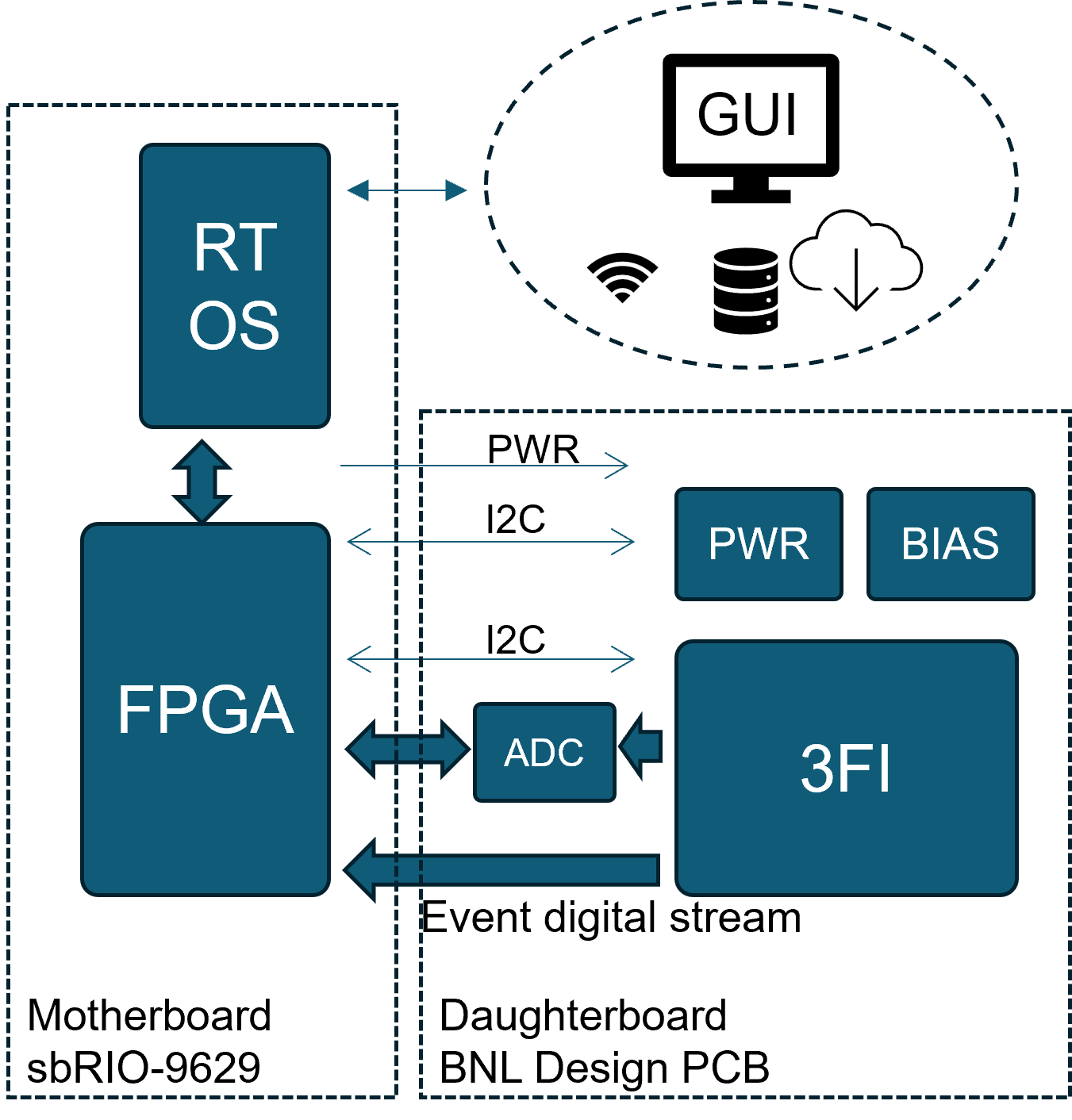}
\end{center}
\caption{Simplified block diagram of a 3FI-based detector.}
\label{fig:DetectorBlockDiagram}
\end{figure}

To facilitate the evaluation of the 3FI ASIC, a fully integrated detector system was developed, supporting both autonomous and supervised operational modes. In autonomous mode, the system can be controlled via TCP/IP commands, allowing for user-defined scripting and seamless integration with existing infrastructures. In supervised mode, a graphical user interface (GUI) provides interactive control for configuration and monitoring purposes. The system is designed for versatility, featuring Peltier-based thermal stabilization to maintain optimal operating temperatures under varying environmental conditions. Additionally, it includes battery-powered and wireless communication capabilities, enhancing its portability and suitability for diverse deployment scenarios. A simplified block diagram of the detector is shown in Fig.~\ref{fig:DetectorBlockDiagram}. The system integrates two main components: an off-the-shelf high-performance embedded controller (sbRIO-9629) and a custom in-house-designed daughterboard. The sbRIO-9629 includes an Artix-7 FPGA and an Intel Atom~E3845 quad-core 1.91~GHz processor, both fully programmable through software.

The presented approach for the architecture of the system results in minimal complexity of the daughterboard, allowing to focus on evaluation of the 3FI ASIC rather than assigning efforts to embedded system development. The daughterboard includes the 3FI ASIC bump-bonded to a 320~\textmu{m} thick silicon sensor, programmable voltage regulators, and both Digital-to-Analog and Analog-to-Digital Converters (DACs and ADCs) for external biasing and monitoring. An AD9649 ADC (14-bit, up to 65~Msps) enables analog signal digitization within a range of +-1V.

The daughterboard interfaces with the embedded controller via a high-density Searay connector, which carries both power and digital input-output signals. The FPGA is configured to perform the following functions:

\begin{itemize}
    \item An I2C interface state machine with configurable clock and data lengths, used for slow control of the 3FI ASIC and auxiliary system components (e.g., power supply, DACs/ADCs).
    \item High-speed transceivers with independently programmable regional clocks, bit-slip correction, and latching delay, ensuring precise synchronization of the 14-bit digital data stream containing event pixel addresses.
    \item Interface logic to the AD9649 ADC with reprogrammable sampling frequencies up to 65~MHz, automatic bit-slip correction, and data delay tuning for accurate digitization of the analog waveform.
    \item A digital stream filtering block that detects valid events by identifying the transition from synchronization patterns to actual event data, reducing the data throughput between FPGA and CPU.
\end{itemize}

The FPGA is managed by software running on a custom Linux Real-Time Operating System (RTOS), which handles key tasks such as FPGA configuration and monitoring, pixel address and ADC data acquisition via Direct Memory Access (DMA), data storage, and an embedded GUI for interactive control. Ethernet connectivity (both wired and wireless) is supported for remote operation and integration with synchrotron infrastructure.

\begin{figure}[ht]
\begin{center}
\includegraphics[width=0.5\textwidth]{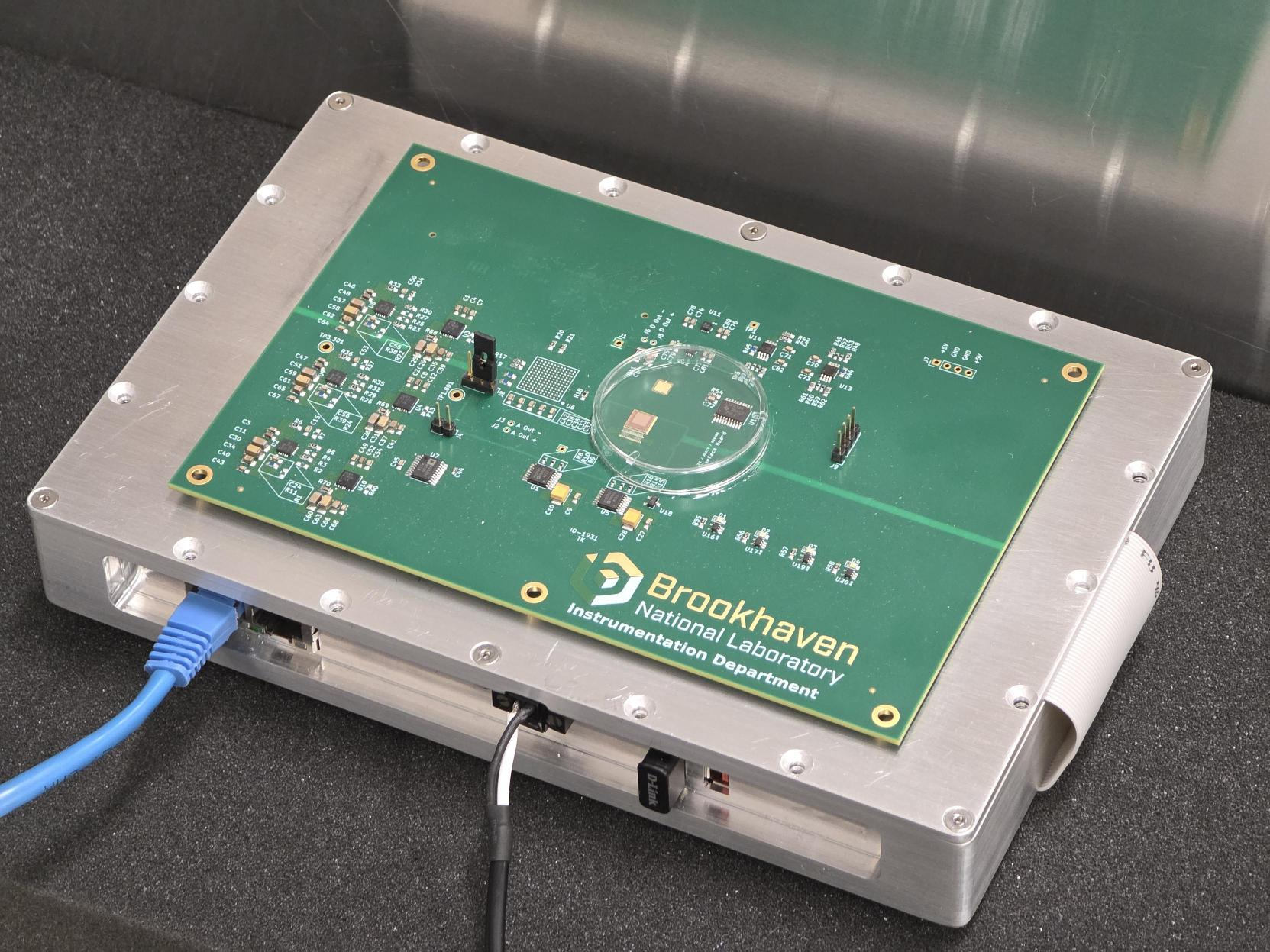}
\end{center}
\caption{Photograph of the detector with the 3FI ROIC - before sensor assembly}
\label{fig:DetPhoto}
\end{figure}

The sbRIO controller is housed within a metal enclosure that provides both passive heat dissipation and electromagnetic interference (EMI) shielding. A connector on the enclosure allows the 3FI daughterboard to be attached externally. A Peltier module is placed between the daughterboard and the enclosure to actively cool the sensor, maintaining temperatures below 0~°C. A photograph of the complete detector is shown in Fig.~\ref{fig:DetPhoto}.

The final system is a fully autonomous, single-photon energy-resolving detector that supports both local and remote operation. The present features include wireless connectivity, real-time data storage, and live data preview. The detector was successfully deployed for its initial evaluation at the beamline 17-BM of NSLS-II (\url{https://www.bnl.gov/nsls2/beamlines/beamline.php?r=17-BM}).

\section{Measurements}\label{Sec:Measurements}

\begin{figure}[ht]
\begin{center}
\includegraphics[width=0.7\textwidth]{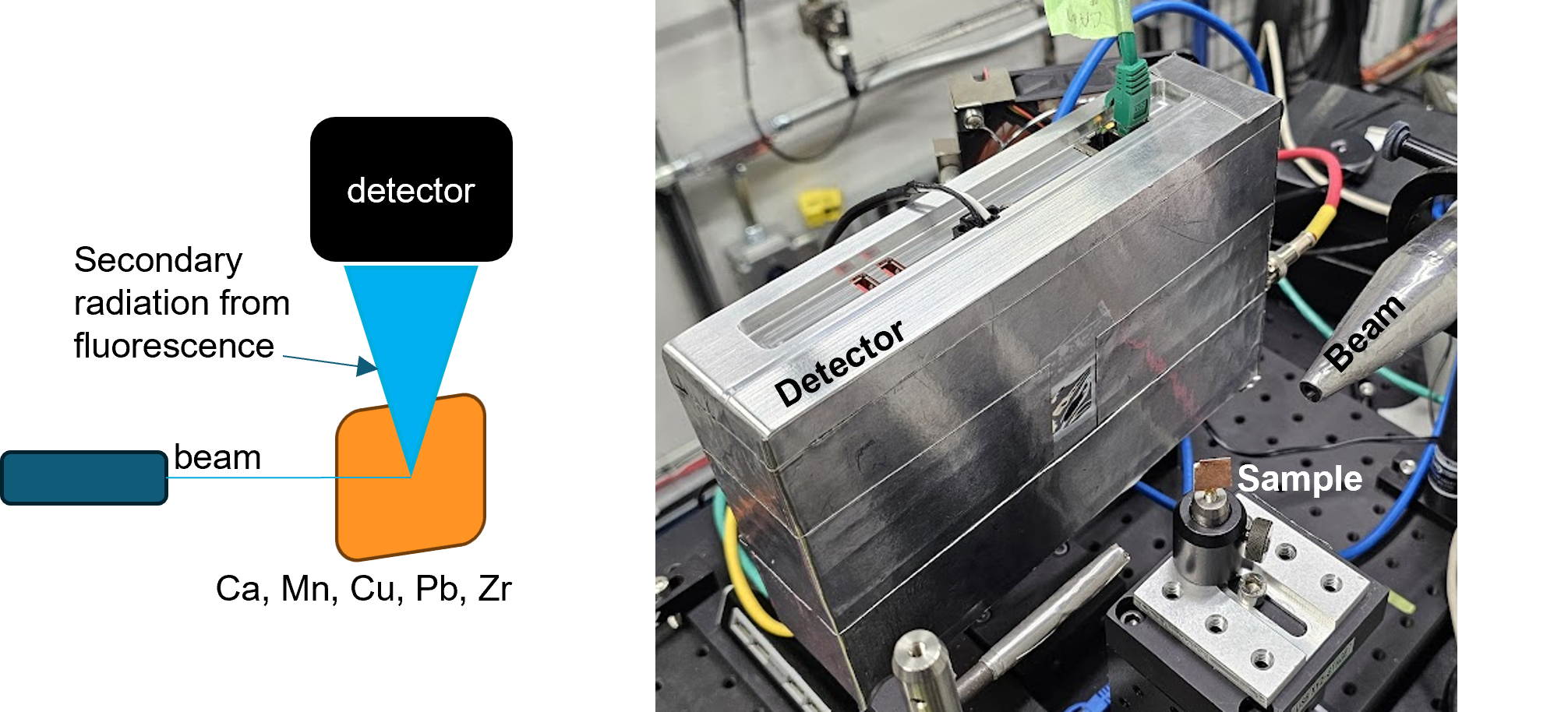}
\end{center}
\caption{Measurement setup: schematic (left) and a setup photo (right) of fluorescence radiation from thin foils of Ca, Mn, Cu, Pb and Zr}
\label{fig:MeasSetup}
\end{figure}

 Comprehensive functional testing verified that both in-pixel and global configuration parameters precisely modulate the operational behavior of the 3FI ASIC, aligning with the intended design specifications. In particular, the ability to trim voltage offset spread at the discriminator input was verified. Quantitative assessment of the 3FI's spectrometric performance was carried within the scenario conceptually shown in Fig.~\ref{fig:MeasSetup}.

\subsection{3FI Functionality Evaluation}

The validation of the detector began with a series of functional tests, defined here as those, that confirm basic power-up, configuration, data–framing and event–handling behavior of the integrated system. The 3FI front-end ASIC dissipates $\sim\!500$\,mW and, once the reference clock is present, immediately transmits a synchronization word that establishes the sampling phase for the downstream FPGA. Correct bi-directional communication was verified through the I\textsuperscript{2}C slow-control bus and through all dedicated control lines (supply rails, sensor-bias nodes and global threshold registers). A valid discriminator response, generated whenever an input pulse crossed the DC baseline, produced time-aligned digital addresses and analogue samples that were captured without skew in the FPGA.

Although 3FI is inherently event-driven, the EDWARD provides a diagnostic ``forced-read'' sequence that activates every pixel in turn~\cite{Gorni_2024}. With the address stream and ADC samples thus aligned, a full-frame image of the $32\times32$ pixel plane was reconstructed, confirming channel ordering and analogue-path integrity.

The detector’s maximum sustainable photon flux is limited by the event-acknowledgement bandwidth. At room temperature the external read-out chain supports 20\,MHz; active cooling extends this up to 40\,MHz. For the
measurements reported here the acknowledge rate was limited to 8\,MHz, allowing 40\,MHz ADC to oversample the latched analogue value five times. Averaging these samples improved the per-event energy estimate while remaining consistent with the zero-dead-time architecture.

\subsubsection{Offset Trimming}

Following the initial verification of the proper 3FI's functionality, the next step was to evaluate the offset trimming capability. Since the discriminator’s DC input level is not directly accessible, a dedicated approach was required. One of the chip's debugging modes enables reading analog values from each pixel's sample-and-hold (S\&H) circuit. By acquiring these values across the full range of trimming DAC settings, it is possible to reconstruct the DC level response for each pixel.

\begin{figure}[ht]
\begin{center}
\includegraphics[width=0.8\textwidth]{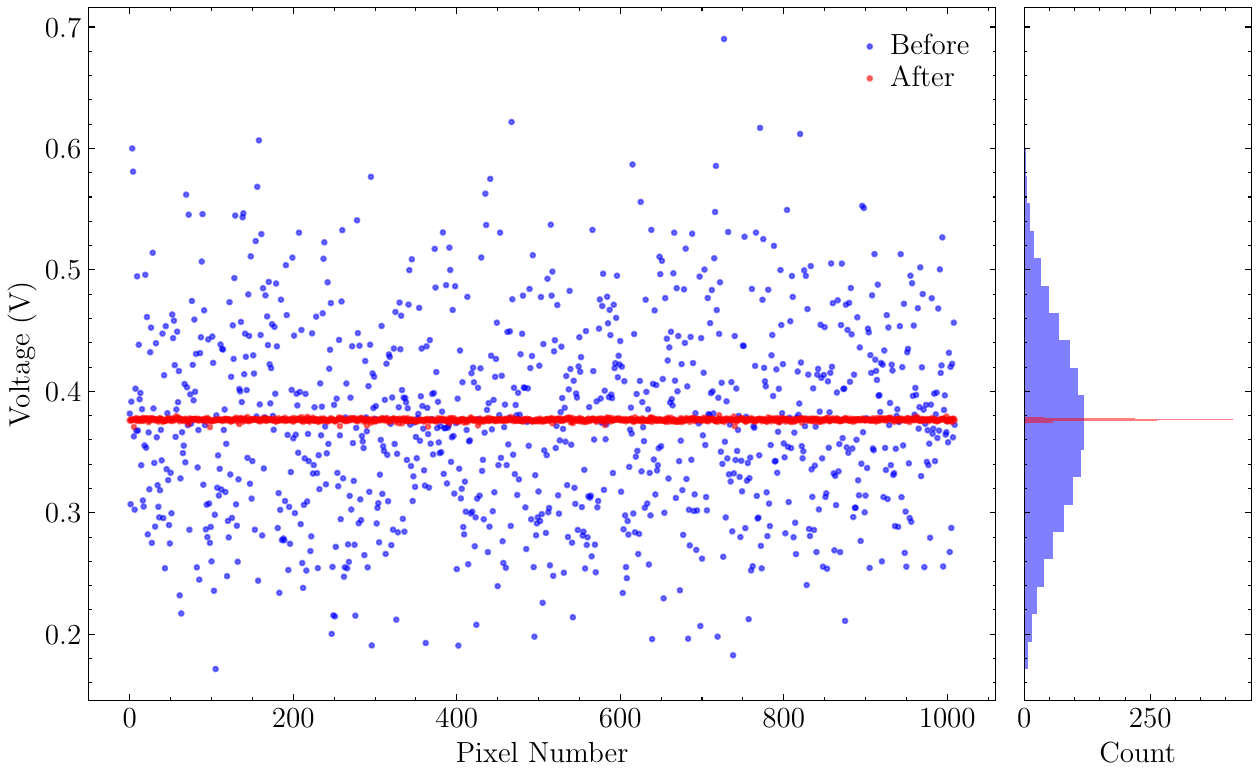}
\end{center}
\caption{DC voltage measured with uniform trim DAC values (a) and after applying the trimming procedure (b). Left: raw data; right: corresponding histogram.}
\label{fig:Trimming}
\end{figure}

Using this data, the optimal DAC settings for minimizing the pixel-to-pixel offset spread can be determined. This method provides a straightforward way to verify trimming performance. Results of this calibration are shown in Fig.~\ref{fig:Trimming}. The standard deviation of DC levels for the same value placed inside the trimming DAC was 79\,mV, and it was reduced to 0.98\,mV after correction, which is consistent with design assumptions and simulation results.

While in binary imaging systems (e.g. Eiger, Medipix), precise setting of the threshold value to the half of the measured energy is essential~\cite{Zhang2018}, in spectrometric systems such as 3FI, the primary role of the threshold is to detect photon events. Therefore, the desired trim DAC setting for each channel is the value that suppresses noise-triggered events while remaining sensitive to photon hits.

To enhance trimming performance under demanding experimental time constraints, an alternative, significantly faster calibration method was developed. While similar techniques have been implemented in photon-counting systems, this approach was specifically designed to accommodate the unique requirements of event-driven readout architectures.

In this calibration approach, the global threshold is fixed at a target value, and all trim DACs are initially set to zero. This configuration places the pixel baselines significantly below the threshold, resulting in non-present noise-triggered event rates. The trim DAC value for each pixel is then incrementally increased while monitoring the average event rate over a fixed time window. During the procedure, each channel crosses the value of predefined threshold generating significant number of events, which then fades with incrementing the trim DAC value further. The final trim DAC setting is identified as the value just above the threshold, at which noise-induced events are effectively suppressed. This method allows for efficient and scalable threshold equalization across the pixel matrix that can execute within several seconds.

This iterative procedure is embedded in the detector control system and can be executed automatically, ensuring fast and reliable offset calibration whenever needed.

\subsubsection{Spectrometric Performance}

\begin{figure}[ht]
\begin{center}
\includegraphics[width=0.6\textwidth]{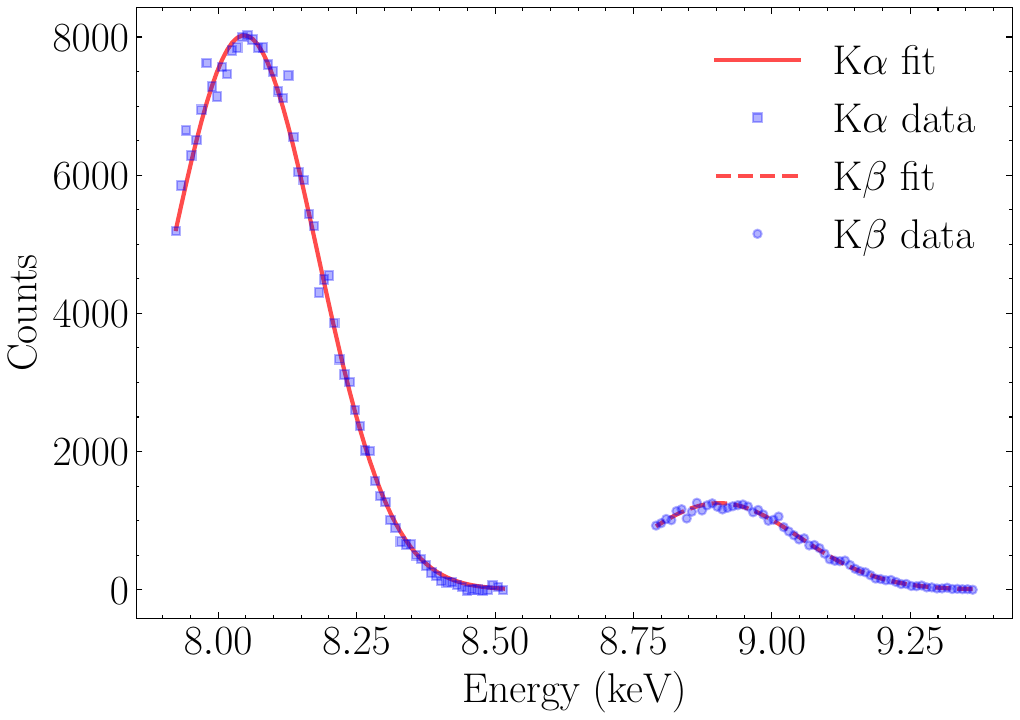}
\end{center}
\caption{Fitting procedure for Cu K\textsubscript{\textalpha} and K\textsubscript{\textbeta} energy peaks.}
\label{fig:Fitting}
\end{figure}

The setup shown in Fig.~\ref{fig:MeasSetup} was employed to evaluate the spectrometric performance of the 3FI-based detector. Operating in charge-sharing compensation mode, the single pixel has latched it's own pulse amplitude and also, in the same time, it has latched values from neighboring channels, enabling the characterization of both the baseline distribution in adjacent pixels and the energy-peak distribution in the active pixel. The pixel selected for detailed analysis was located centrally within a region of intact sensor connections. Energy resolution was assessed by acquiring fluorescence spectra from materials with well-defined X-ray emission lines. Noise estimation was performed by fitting the right-hand side of each energy peak with a Gaussian function, as the left side typically contains background contributions~\cite{BLAND19981225}. An example of this fitting procedure, applied to the Cu K\textsubscript{\textalpha} and K\textsubscript{\textbeta} lines, is presented in Fig.~\ref{fig:Fitting}.

The target materials produced characteristic K\textsubscript{\textalpha} and K\textsubscript{\textbeta} fluorescence lines, with emission energies as follows:

\begin{itemize}
    \item \textbf{Calcium (Ca)}: K\textsubscript{\textalpha} at 3.691\,keV, K\textsubscript{\textbeta} at 4.012\,keV
    \item \textbf{Manganese (Mn)}: K\textsubscript{\textalpha} at 5.898\,keV, K\textsubscript{\textbeta} at 6.490\,keV
    \item \textbf{Copper (Cu)}: K\textsubscript{\textalpha} at 8.047\,keV, K\textsubscript{\textbeta} at 8.905\,keV
    \item \textbf{Lead (Pb)}: L\textsubscript{\textalpha} at 10.551\,keV, L\textsubscript{\textbeta} at 12.667\,keV
    \item \textbf{Zirconium (Zr)}: K\textsubscript{\textalpha} at 15.775\,keV, K\textsubscript{\textbeta} at 17.667\,keV
\end{itemize}

\begin{figure}[ht]
\begin{center}
\includegraphics[width=0.8\textwidth]{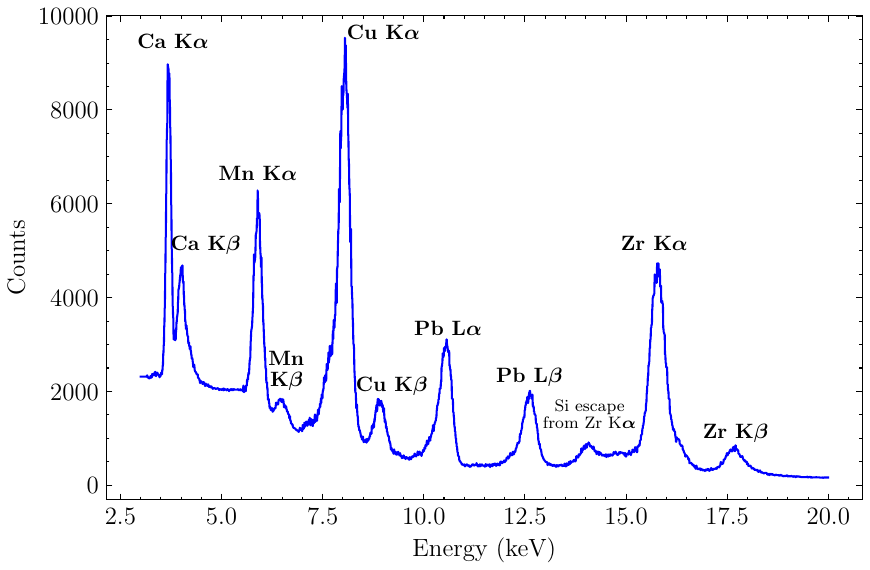}
\end{center}
\caption{Combined energy spectrum from fluorescence targets.}
\label{fig:SpectrumCombined}
\end{figure}

To ensure statistical relevance, measurement durations were adjusted to collect more than two million events per target. The resulting spectra were normalized to background event rates and combined into a single spectrogram, shown in Fig.~\ref{fig:SpectrumCombined}.

The energy resolution for triggered pixel only (without charge-sharing compensation) and for the Ca K\textsubscript{\textalpha} line was measured at 138\,eV full width at half maximum, corresponding to an equivalent noise charge of 16\,e\textsuperscript{-} rms. The FWHM resolutions measured for other K\textsubscript{\textalpha} lines (and L\textsubscript{\textalpha} for Pb) were as follows: Mn at 249\,eV, Cu at 308\,eV, Pb at 339\,eV and Zr at 469\,eV. In all cases, clear peak separation was observed, demonstrating the detector's low-noise performance and excellent spectrometric capabilities. No significant nonlinearity was detected across the evaluated energy range, confirming the suitability of the 3FI system for color X-ray imaging applications.

For each photon event detected in the central pixel, the readout also included values from adjacent pixels. These neighboring channels predominantly contained baseline values, as the probability of charge sharing with a particular neighboring pixel, especially corner pixels, was low. The distribution of these baseline measurements defines the baseline noise level, determined to be approximately 120\,eV FWHM, consistent with simulation results and measured energy peaks.

%

\section{Conclusions}

This work presents the design, implementation, and validation of the 3FI ASIC, a novel event-driven readout chip developed at Brookhaven National Laboratory for high-speed, energy-resolved X-ray fluorescence imaging. The 3FI features a 32~×~32 array of 100\,µm~×~100\,µm pixels, each equipped with a charge-sensitive amplifier, shaping filter, peak detector, and local sample-and-hold circuitry. Fabricated in a 65\,nm CMOS process, the ASIC incorporates the EDWARD event-driven architecture, enabling fast, per-pixel spectral acquisition with frameless data processing.

Integrated into a compact, fully autonomous detector system with embedded control, wireless connectivity, and real-time data handling, the 3FI was successfully evaluated at beamline 17-BM (XFP) at NSLS-II. The detector demonstrated energy resolutions as low as 138\,eV at 3.7\,keV and 308\,eV at 8.0\,keV, with clear spectral separation across a wide energy range. Neighboring pixels baseline noise distribution measured at 120\,eV confirms the low-noise performance. Automated offset trimming and a low-noise front-end support stable operation and robust spectroscopic performance.

Designed with portability and integration in mind, the 3FI detector is easily deployable at synchrotron facilities. It can interface directly with beamline control systems and is capable of controlling external hardware such as XYZ stages or other auxiliary instrumentation, supporting advanced experimental automation.

By combining deep submicron CMOS technology with a high-throughput, asynchronous readout paradigm, this work establishes 3FI as a promising platform for next-generation, full-field X-ray fluorescence imaging, particularly in applications requiring rapid, high-resolution elemental mapping of complex biological and environmental samples.


\begin{funding}
This work has been co-authored by employees of Brookhaven Science Associates LLC under Contract no. DE-SC0012704 with the U.S. Department of Energy. 
\end{funding}

\ConflictsOfInterest{There are no conflicts of interest.
}


\bibliography{biblio} 

\end{document}